# The Fourth Dimension

Eugen Schweitzer
aithon@gmx.de

## Abstract

Plato indicated at different passages of his dialogues deep mathematically based physical insights. Regrettably the readers overlooked the respective statements or being full of philology they utterly did not understand those hints. Respectable translators misinterpreted such statements and therefore Plato's respective remarks had not been recognized as substantial knowledge. Furthermore, Plato often supplemented such basic remarks by dispersed allusions diffusely veiled and often ironically hidden somewhere in his dialogues by inconspicuous double meanings. However, this mode of intentionally coded discrete communication had generally not been understood because such irony is not to everyone's taste. However, the attempts to reconstruct Plato's systematic on the base of admittedly individually interpreted double meanings lead to a conclusive mathematical-physical cyclical system of dimensions. Additionally it was possible to assign Plato's system of philosophical ideas analogously to this cyclical system. Plato took the verifiability of the mathematical-physical results as proof of the systematic of his ideas and finally as proof of his ethical creed, the unconditional trust in the all surmounting Good.

## Plato's Fourth Dimension

In a passage which deals with the essential difference between understanding and reasoning, between knowledge and cognition Plato refers to the fourth dimension as speeds:

> Of course we have to divide the art of measurement into two parts just as we said. On the one hand we put all the arts which measure number (arithmon), lengths (mêkê), depths (bathê), breadths (platê), and speeds (tachutêtas) in relation to what is opposed to them, on the other there are all those that measure in relation to what is moderate, suitable, opportune, needful and all together that is settled as mean value of the extremes. (*Statesman* 284e)

The interesting first part of the main sentence here implies a conceptual mixture of terms: number (sg) as a mere mathematical term, length(s), depth(s), and breadth(s) as geometrical terms (pl) of dimensions, and speed(s) as a physical term of motions. It is known that Plato considered the first unit of dimensions a number and not a point, contrasting today's terminology.

Want of inquisitiveness and preconceived evaluation might have led the principal translator of Plato's work into English to find a putative contradiction in terms in this sequence of terms. In addition, he seemed to be unaware of any coherence between mathematics, geometry, and physics. This might have led to his well intentioned arbitrary translation of the plural form of the Greek word 'tachutés' as the English singular form 'thickness' (Fowler H.N.) instead of correctly using 'speeds' (Autenrieth) ('quickness', 'swiftness' in LSJ and ML).

In another passage (*Statesman* 299e), the same translator interprets the same term as 'problem of motion'. These mistranslations have adulterated the meaning of the originals to nonsense out of all reason. Yet, despite the fact that there are nearly correct translations e. g. into the German as the singular form 'Geschwindigkeit', none of the countless interpreters noticed the comprehensive and unifying meaning in this combination of mathematical, geometrical and physical terms. Gaiser, a philologist who researched extensively about Plato's thinking about dimensions not even mentioned those quotations regarding this sequence of dimensions (Gaiser. 107–115).

However, this mixed sequence of terms shows that Plato had a unifying point of view. As the mathematical term set up a geometrical dimensional sequence followed by a physical term, jointly they can be looked at as a general link to exponentiation, whereas the mathematical and geometrical terms also acquire a physical quality. Accordingly Plato's sequence of dimensions meets the general geometrical sequence of point, line, plane, and space, and the fifth term in this kind of strange order, being the term 'speeds', can be viewed as Plato's definition of a physical fourth dimension. Also space(s) as third dimension applies accordingly to a mathematical and geometrical, as well as to a physical meaning. Altogether, these five terms of consecutive values of dimensions stand for that type of application of mathematics (*mathesis intensortim*, cf. Kant Prolegomena II 24) to the science of nature.

**Dimensions as Rectangular Motions**

For an accessible and reviewable explanation of the physical quality of Plato's forth dimension we have to recognize the unifying structure of mathematical powers and their physical equivalences. Plato defined the zero dimension, geometrically represented by a point, as 'number', and the dimension to the power of one is geometrically regarded as line(s), whereas in physics it is generalized as length(s). We can define $d^1$ (d for distance) as the result of a movement of a point $d^0$, expressed as $d^0 \cdot d^1 = d^1$.

The abstract mathematical term of a dimension (power) masks this notional movement that is pointing indirectly to an undefined time during the passing of an undefined distance. The movement $d^1$ of a line $d^1$ in any right angle to its direction forms a plane $d^2$, expressed as $d^1 \cdot d^1 = d^2$. A further motion $d^1$ rectangular to the extension of a plane $d^2$ gives a space $d^3$, expressed as $d^2 \cdot d^1 = d^3$ . Space seems due to those threefold movement to have beside his abstract mathematical definition as $d^3$ an own concrete physical quality to be considered.

Defining the fourth dimension $d^4$ as rectangular motion $d^1$ of a space $d^3$ (respectively volume, body or matter) we get $d^3 \cdot d^1 = d^4$ or expressed otherwise: If we regard this fourth dimension figuratively as the motion of a body it meets the notion speed as quantity of the quality 'velocity'.

It has to be mentioned that according to the abstract definitions of the geometrical qualities of dimensions, we use corresponding expressions when we talk about their physical quantities: line –length; plane – area; space – space and volume; velocity – speed. For example, speed does not contain the element of direction that velocity has and determines as magnitude component the quantity of velocity. However, both definitions contain the notion of movement.

## ‘Unit circle in the Complex Plane’

As we explained, the step from one dimension to the other is a *rectangular movement* of the foregoing dimension. Four of those *rectangular rotation* leads to the 360° of a circle and to an association with the ‘Unit circle in the Complex Plane’ (Figure 1).

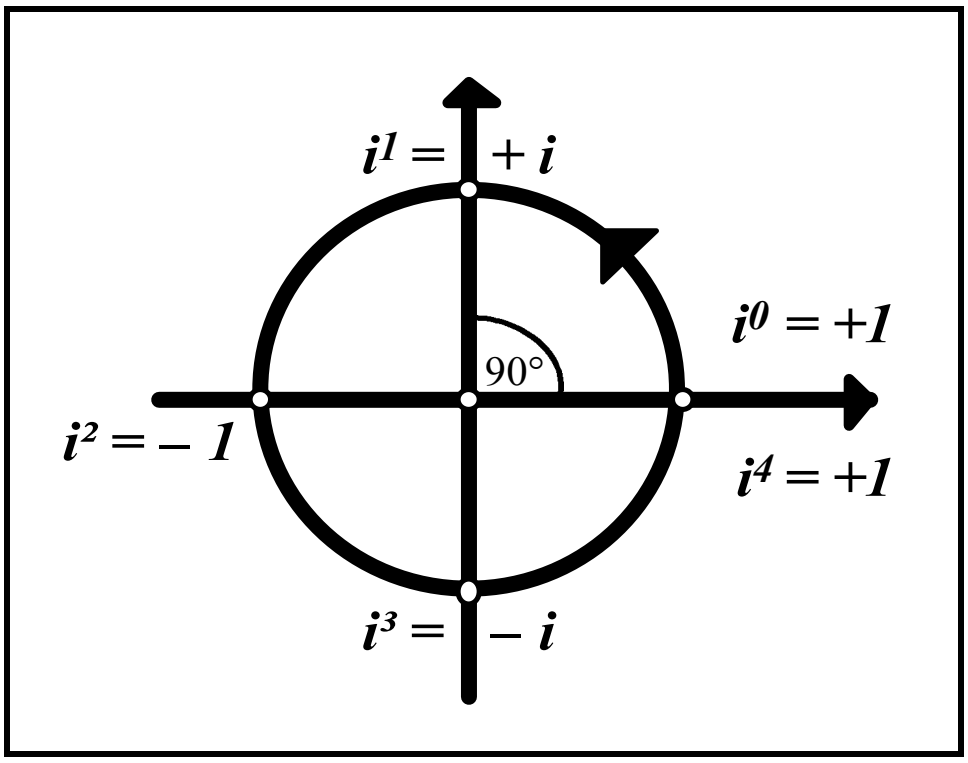

Fig. 1: ‘Unit Circle in the Complex Plane’

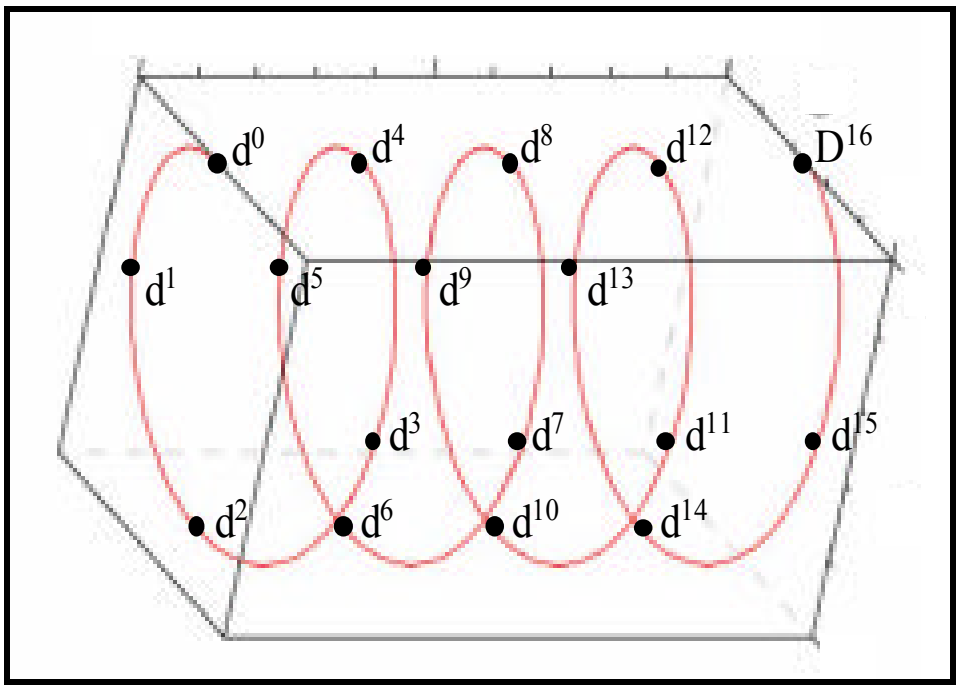

Fig. 2: Spiral Sequence of Dimen-

After the first cyclical turnabout the next circle starts as a new ‘lower cyclic’. And four of such cycles form an upper cyclic of 16 dimensions and so on. We might insert for *dimension* the shortage *d* according to the complex term *i* at the ‘Unit Circle in the Complex Plane’. This Unit Circle might be regarded as the plain projection of an infinite simple spiral of dimensions (Figure 2).

For a better survey we can transpose the rectangular steps of the respective quadrants of a first circle into the first column of a table. The next turn of the circle will be represented by the second column, and four of such columns constitute an upper cycle with 16 dimensions (Figures 3, 4, 5, 6).

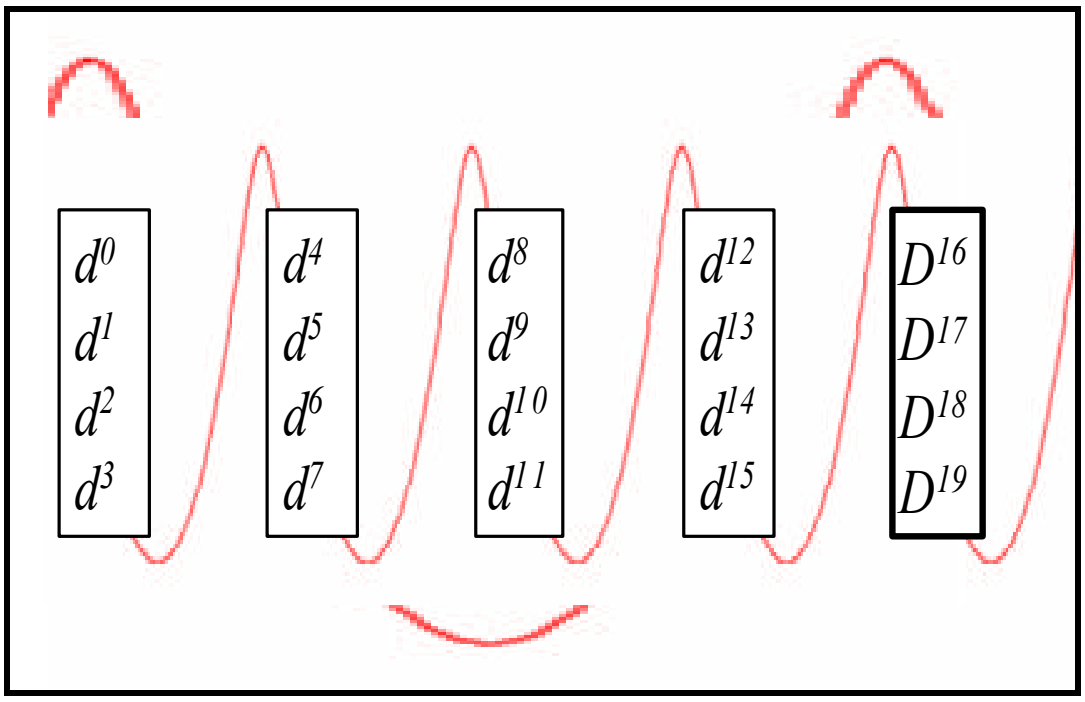

Fig. 3: Summing up of Dimensions as Pre-stage of a Matrix

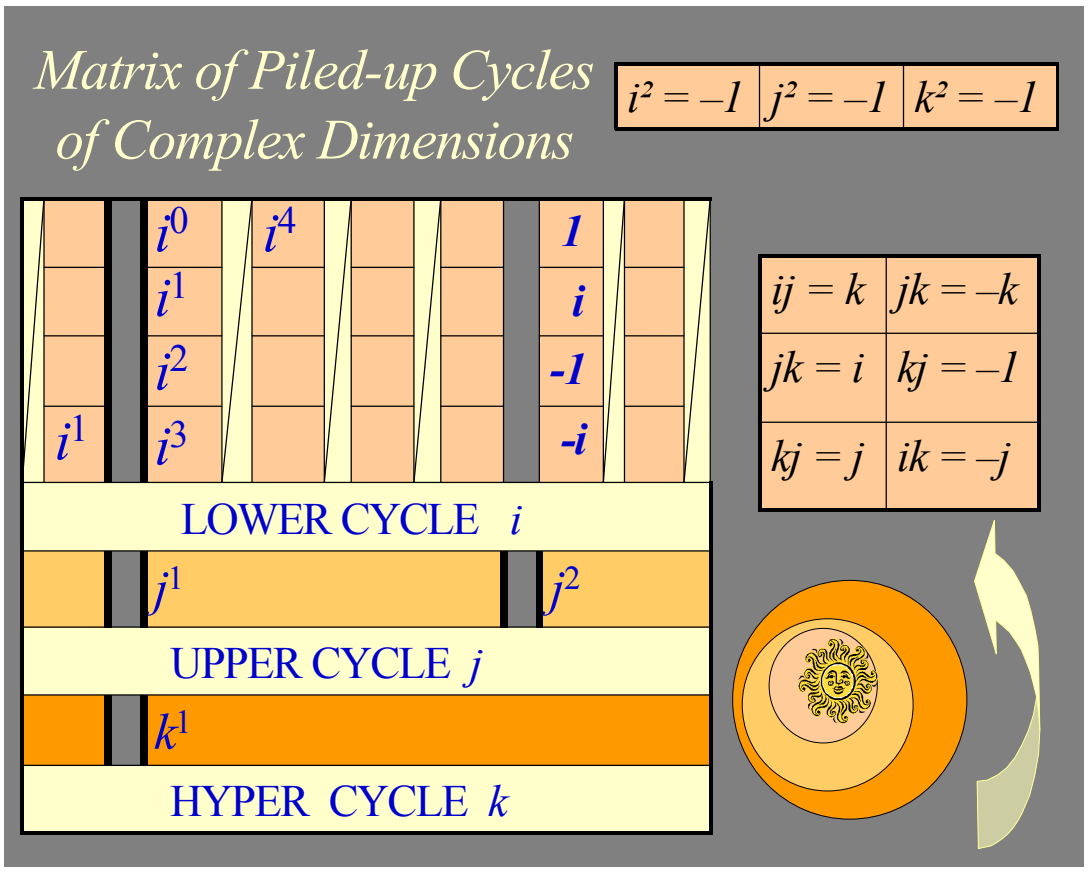

Fig. 4: Matrix of Piled-up Cycles of Complex Dimensions of Lower, Upper, and Hyper Cycles,

With respect to the cyclical principle the further upper and hyper cyclices lead via Hamilton's quaternions to a method of analysis even beyond octaves.

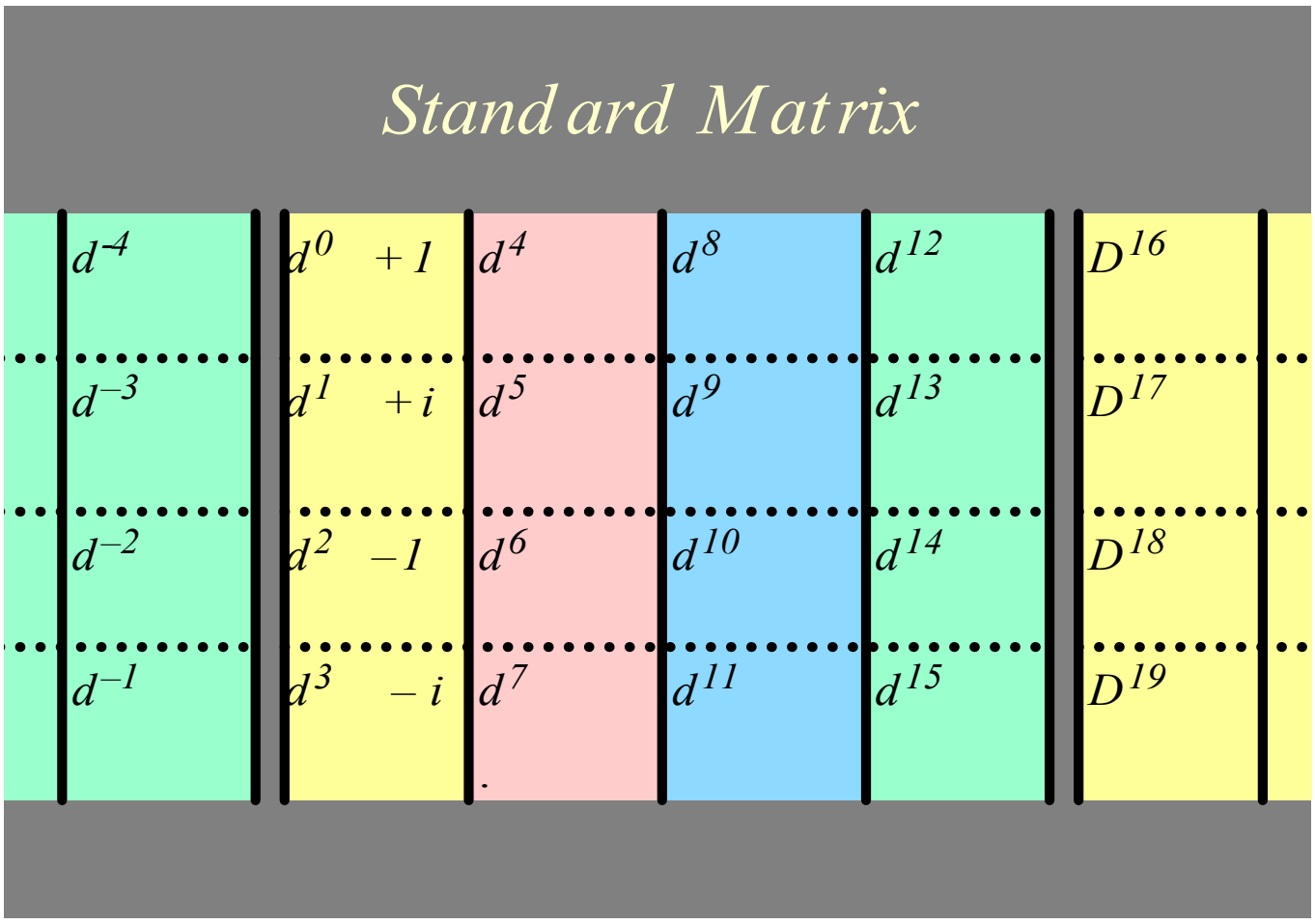

Fig. 5: Unified Pattern of a Matrix of Piled-up Values

## Visualisation of Plato's Physical Dimensions

Plato's sequence of physical dimensions can be projected to the 'Unit Circle in the Complex Plane' and the respective matrix might serve for a better understanding of the continuation beyond the fourth dimension. Plato's physical dimensional system is based on a self motioned fundamental force which enables the step from $d^0$ to $d^1$ and so on.

| | | | |
|---|---|---|---|
| $d^0$ $+1$ | $d^4$ | $d^8$ | $d^{12}$ |
| $d^1$ $+i$ | $d^5$ | $d^9$ | $d^{13}$ |
| $d^2$ $-1$ | $d^6$ | $d^{10}$ | $d^{14}$ |
| $d^3$ $-i$ | $d^7$ | $d^{11}$ | $d^{15}$ |

Fig. 6: Standard Matrix Pattern

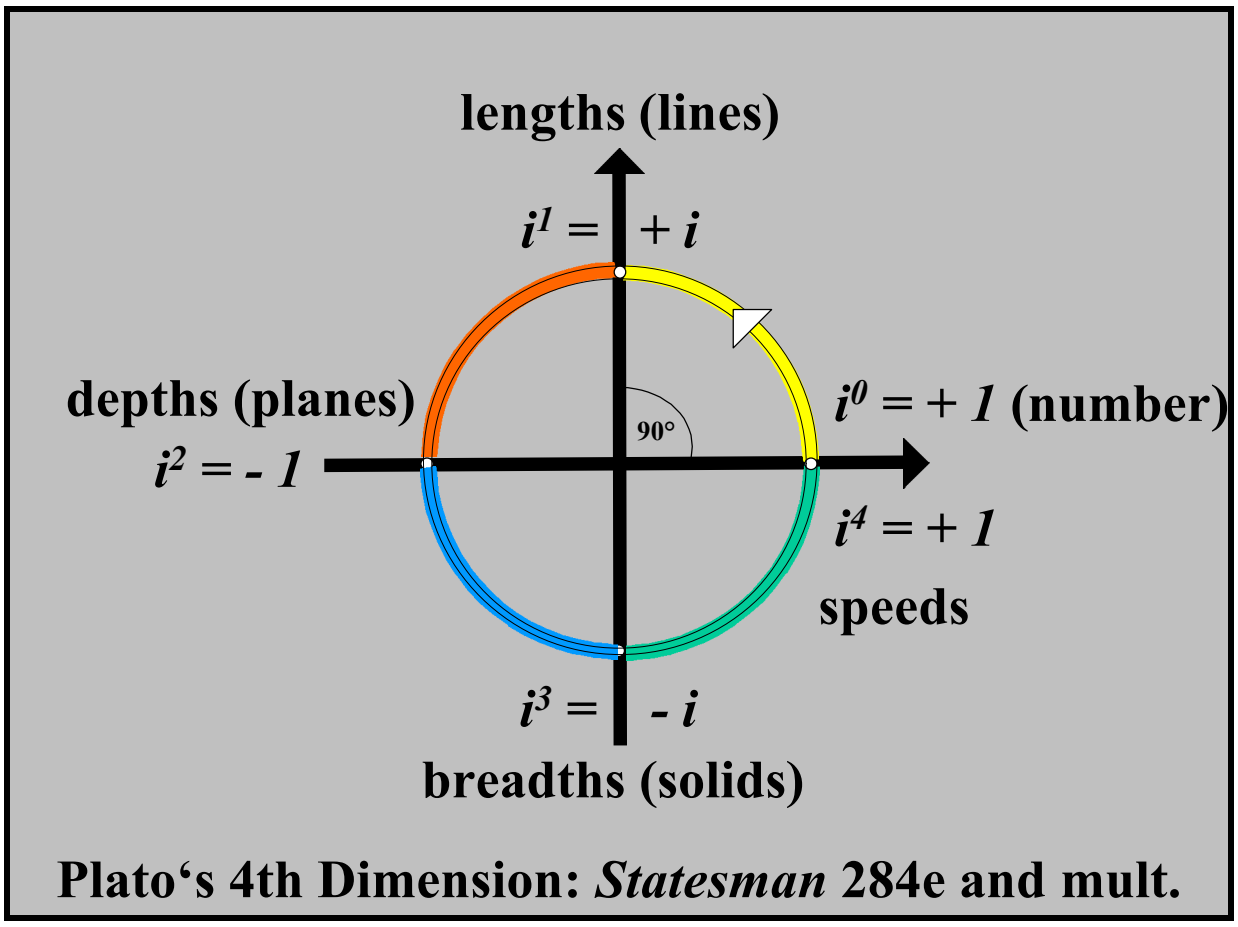

Fig. 7: Plato’s Sequence of Dimensions adopted to the ‘Unit Circle in the Complex Plane’

*Speed as Moved Space*

| | | | |
|---|---|---|---|
| $d^0$ +1 point selfmotion | $d^4$ selfmotioned space = veloc. speed v | $d^8$ | $d^{12}$ |
| $d^1$ $+i$ selfmotioned point = line length $s$ | $d^5$ | $d^9$ | $d^{13}$ |
| $d^2$ −1 selfmotioned line = plane area A | $d^6$ | $d^{10}$ | $d^{14}$ |
| $d^3$ $-i$ selfmotioned plane = space volume V | $d^7$ | $d^{11}$ | $d^{15}$ |

Fig. 8: Speeds as Moved Spaces

The projection of this spiral as ‘Unit Circle in the Complex Plane’ is exactly related to Plato’s sequence of dimensions (Figure 7), which also can be demonstrated in the corresponding matrix (Figure 8). The consequence regarding time and its interdependence can be exemplified in an extended matrix of respective dimensions (Figure 9).

*4th Dimension*

| | | | | |
|---|---|---|---|---|
| $d^{-4}$ | **$d^0$ +1** point | $d^4$ **speed** | $d^8$ | $d^{12}$ |
| **$d^{-3}$ time** | $d^1$ **$+i$ line** | $d^5$ | $d^9$ | $d^{13}$ |
| $d^{-2}$ | $d^2$ **−1** plane | $d^6$ | $d^{10}$ | $d^{14}$ |
| $d^{-1}$ | **$d^3$ $-i$** space | $d^7$ | $d^{11}$ | $d^{15}$ |

**v = $d^4$ = path / time = $d^1$ / $d^{-3}$**
**t = path / speed = $d^1$ / $d^4$ = $d^{-3}$**

Fig. 9: The Relation of Speed and Time

As we define speed as a distance covered in a past time, we usually express speed by the formula ‘speed is equal path per time’ or v = s / t. Transcribed into those dimensions it is: $d^4 = d^1$ / time, we can solve this equation after the time and we get the result: time = $d^1 / d^4 = d^{-3}$, which means that the comprehensive notion ‘time’, related to a three-dimensional space, be the inversion of the global notion of this three-dimensional space: $d^{-3} = 1 / d^3$, or *TIME* = 1 / *SPACE* (Figure 10).

This definition of a three-dimensional-space-related time $d^{-3}$ as inversion of such space meets a surprising coherence between the general notion of ‘spacetime’. Simplified speaking, this coherence could mean that the passing three-dimensional time corresponds with the universal three-dimensional expanding space being a three dimensional universe. It is now a new paradigm that the physical dimensions of time and space are like one coin with two inverse sides.

The continuous expansion of the universe caused by the underlying self-motion measures time like the frequency of calibrated clockwork: 1 / time = $d^3$ . “He (God) made an eternal image, moving according to number, even that which we have named Time” (*Timaeus* 37d).

With a negative exponent as $d^{-3}$ time is quasi one-directional backward orientated and therefore irreversible. Summing up, this examination shows that time is determined by its position in the cyclic system being the inversion of the summing up of all quasi third dimensions of space of all revolving cycles. Inversion of a complex dimension means it’s mirroring on the real ordinate into the unreal values. This mirrored image of space $d^3$ as the dimension of time $d^{-3}$ being with its negative exponent an inversion of space is an unreal complex value, a mere arithmetic value.

*Time as Inversion of Space*

| | | | | | |
|---|---|---|---|---|---|
| $d^{-4}$ | $d^0$ $+1$ **point** | $d^4$ | $d^8$ | $d^{12}$ | $D^{16}$ |
| $d^{-3}$ **time** | $d^1$ $+i$ line | $d^5$ | $d^9$ | $d^{13}$ | $D^{17}$ |
| $d^{-2}$ | $d^2$ $-1$ plane | $d^6$ | $d^{10}$ | $d^{14}$ | $D^{18}$ |
| $d^{-1}$ | $d^3$ $-i$ **space** | $d^7$ | $d^{11}$ | $d^{15}$ | $D^{19}$ |

Fig. 10: The Relation of Time and Space

*The Linear Aspect of Time*

| | | | | | |
|---|---|---|---|---|---|
| $d^{-4}$ | $d^0$ $+1$ point | $d^4$ | $d^8$ | $d^{12}$ | $D^{16}$ |
| $d^{-3}$ **time** | $d^1$ $+i$ **line** | $d^5$ | $d^9$ | $d^{13}$ | $D^{17}$ |
| $d^{-2}$ | $d^2$ $-1$ plane | $d^6$ | $d^{10}$ | $d^{14}$ | $D^{18}$ |
| $d^{-1}$ | $d^3$ $-i$ space | $d^7$ | $d^{11}$ | $d^{15}$ | $D^{19}$ |

Fig. 11: Time and Line are Linear

Time has with reference to the 'Unit circle in the Complex Plane' the cyclic position at $d^{-3}$, which corresponds to the cyclic position of the *linear* dimension of $d^1$. Explicitly, the linear aspect can be seen regarding both, time and line being in the second row of the matrix (Figure 11), even in an agglomerated manifold of cycles. This might explain why we experience the passing of time as rectilinear.

The seemingly stable three-dimensionality of the experienced space is the result of the superposition of all those loci of all third dimensions of all upper cycles. This seemingly matches the experienced space with its ruling imagination of a three-dimensional Euclidian Space. However, this kind of space here holds more than the result of ordering axioms that served as a feasible arrangement for Newton's inertial frame. Here all spaces are contained in each other forming one just seemingly stable universe like a Hausdorff Space. Though, based on its self-motioned origin, this agglomerated space is more comprehensive, it is not static, but dynamically expanding as 'Platonic space' like our observed universe. According to these discoveries time does not really exist, it is just fictional. It can only be measured as the past, as past expansion of space, as an image of a factual difference of self-moved expanding space.

This space-time relation may perhaps correspond in a certain way to the space-time continuum of modern physics. Einstein combined in a visionary synopsis space and time creating as a unifying notion a single construct called the 'spacetime' continuum. Hereby space being three-dimensional, and time regarded as one dimensional, are assigned together to play the role of a fourth dimension. Einstein needed a complex auxiliary mathematical construction to connect those dimensions of time and space to the not comprehendible indistinct and artificial notion of a seemingly four-dimensional construct called 'spacetime'. However, Einstein recognized a interrelation of time and space and the application, of their mathematical coinage, even if unnecessarily complicated and unimaginative, made modern physics more efficient simplifying and unifying a large number of physical theories concerning supergalactical as well as subatomic levels. Now however, it is a mental experiment to dare thinking outside this box and to analyse the cognition that the internalisation of the term velocity meets the definition of the fourth dimension simply better. The acceptance of Plato's insight, that time is just the inversion of an expanding space, simplifies his physics.

## The Interconnection between the Dimensions

Mechanical physics mainly is based on the fundamental notions of mass, length, and time. Those notions can be used to describe any condition of every simple physical situation or process. In classical physics the concepts of those notions are not concerned with each other. E.g. in classical mechanics time is treated as one dimensional, universal and constant and the respective universe has three dimensions of an independent 3-dimensional space containing independent 3-dimensional masses.

However, today's elementary physics takes it for granted that there is some kind of interrelation between those fundamental physical dimensions. In Einstein's relativistic contexts, time cannot be separated from the three dimensions of space, because in his understanding the rate at which time passes depends on an object's velocity relative to the speed of light and also the strength of intense gravitational fields of masses which can slow the passage of time. However, this mutual influence of speed and mass is simply a mere effect and not the reason of the interconnection of time and space. In Plato's physics the factual reason of the interconnection of all physical dimensions is the dimensional configuration of all physical values as result of the self moving expanding basic power leading from one dimension to the next. It looks like the natural increase of potencies, like vivid mathematics with potencies growing from one dimension to the next, inspiring fantasies about fractional powers and their images in nature.

The problem of trying to understand this kind of theory of relativity is the limited imagination of a physical fourth dimension connected with the question of the physical meaning of further dimensions. Mathematically there is no problem to realize an unlimited sequence of powers.

## Mass and Energy

Now we put the question after the dimensional valence of mass, which provisionally had been postponed. As force can be defined as the cause for a change of the state of movement of a body, ‘force’ can be defined as ‘mass multiplied by acceleration’ F = m a. This general definition of force F = m a can be dissolved after mass: m = F / a. (Figure 12). Now we have additionally two unknowns, force and acceleration. The dimensional valence of acceleration can easily be found through its usual definition as change of speed per amount of time. As the mathematical value of speed is $d^4$, the valance of acceleration is accordingly a = v / t or $d^4 / d^{-3} = d^7$. As the cycle of dimensions starts with $d^0$ as beginning of a dimensional development initiated by a self motion, it might be concluded that this first position $d^0$ is a genuine power. Since Plato reserves the value $d^0$ for the minimal original force of self-motion, it implies that to denote a concrete physical force requires starting with at least the valence of the next upper cycle, which is dimension $D^{16}$.

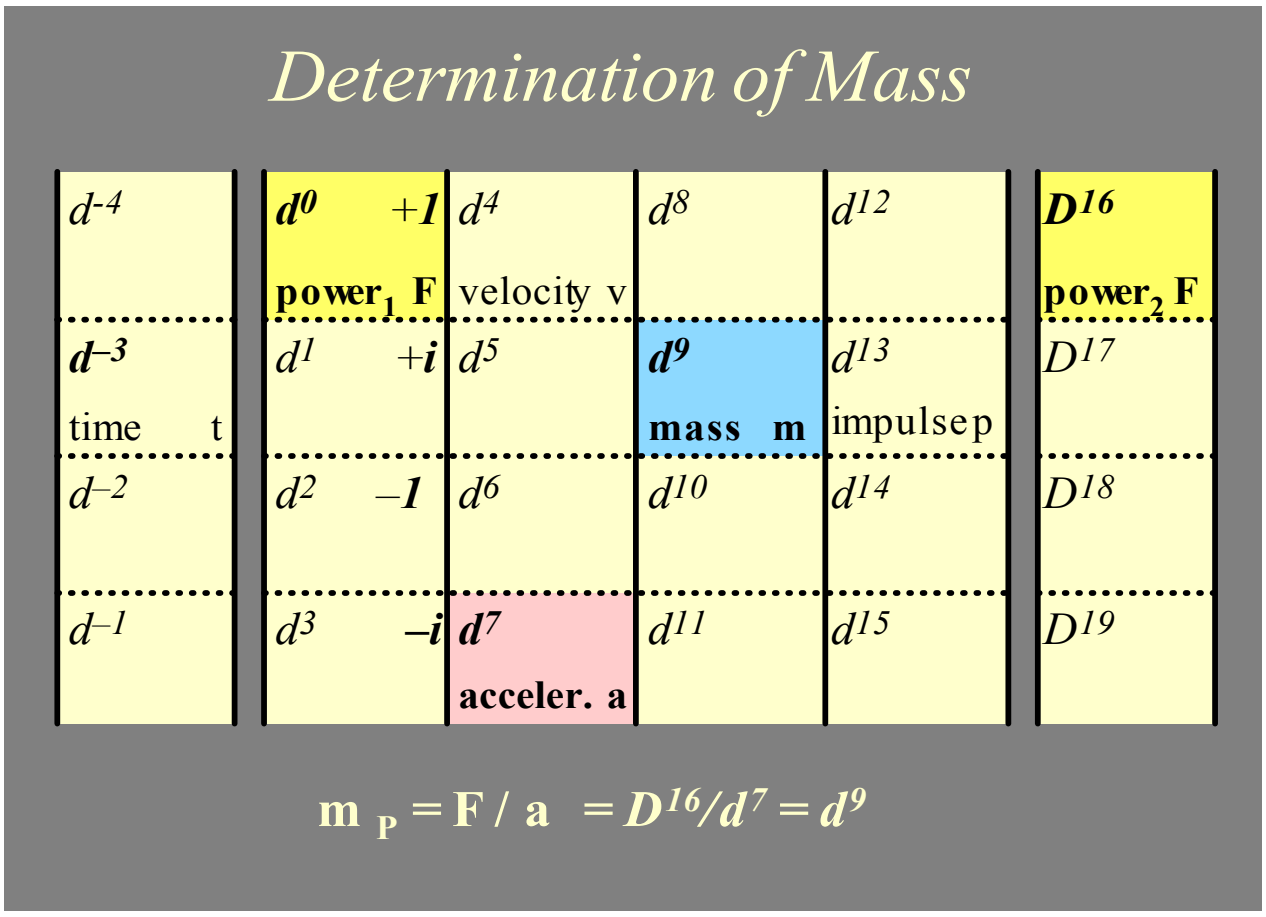

Fig. 12: Dimensional Determination of Mass

As the first position of the second lower cycle is found to be velocity and not per se a genuine power, it might be taken into serious consideration that the physical valence of this $D^{16}$ as starting dimension of the next upper cycle being an image of the paradigm force $d^0$ has to be regarded as a substantial power. Hence, in this new system of interdependent physical dimensions, further types of forces after $d^0$ must have the dimensions $D^{16}$, $D^{32}$, and so forth (Figure 12).

Those dimensional valences of acceleration as $d^7$ and force as $D^{16}$ enables us according to the derivation of today's definition of force to write the definition of mass being m = F / a in dimensions as $D^{16} / d^7 = d^9$. Now the question is how our dimensional valences of force and mass comply with the dimensional system.

The position of forces at $d^0$, $D^{16}$, $D^{32}$ and so forth is a conclusive and continuative result which leads us as next step to the classification of energies within this dimensional system. The usual definition of mechanical energy is force multiplied by path. Expressed in powers of our dimensional system it is $E = d^0 \cdot d^1 = d^1$. The cyclical development also means that in each upper cyclic we find the next form of energy as $E_2 = D^{16} \cdot d^1 = D^{17}$ or $E_3 = D^{32} \cdot d^1 = D^{33}$. In further cycles, the physical values of the respective energy are repeated in respective graduations. This implies multiple forms of energy.

## The Elements

In the cosmology of his *Timaeus* Dialogue, Plato described the inner relationship of elements: "air being to water as fire is to air, and water being to earth as air is to water" (*Timaeus* 32d).

| | | | | | | | | | | |
|---|---|---|---|---|---|---|---|---|---|---|
| FIRE | : | AIR | = | AIR | : | WATER | = | WATER | : EARTH | |
| a | : | f(a) | = | f(a) | : | f ′(a) | = | f ′(a) | : f ′′ (a) | |
| $d^1$ | : | $d^5$ | = | $d^5$ | : | $d^9$ | = | $d^9$ | : $d^{13}$ | $= d^{-4}$ |

This ratio is not a mere *triplasion logos* as Euclid describes it in his 10th Book (a : b = b : c = c : d). Plato's sequence of ratios implies an inner relation of the elements. The mathematical consequence is to regard the elements physically as dimensions with a common base.

In relation to modern physical terms, 'sun' stands for the primordial, self moved 'power' of the Good, and the element 'fire' is the image of the 'sun' and represents 'energy' (Figure 13). Though there is little knowledge about Plato's physical terminology, the physical meaning of his elements correspond with modern definitions: fire = energy, air = aether, water = mass, and earth = impulse.

*Analogies*

| | | | |
|---|---|---|---|
| $d^0$ **+1**<br>$D^{16}$ **SUN**<br>GOOD | $d^4$<br>**STARS**<br>True | $d^8$<br>**MOON**<br>Beautiful | $d^{12}$<br>**EARTH**<br>Moderate |
| $d^1$ $+i$<br>**FIRE**<br>brave | $d^5$<br>**AIR**<br>wise | $d^9$<br>**WATER**<br>just | $d^{13}$ $d^{-3} = t$<br>**EARTH**<br>temperate |
| $d^2$ $-1$ | $d^6$ | $d^{10}$ | $d^{14}$ |
| $d^3$ $-i$ | $d^7$ | $d^{11}$ | $d^{15}$ |

Fig. 13: The Analogy of Elements and Ideas and their Images

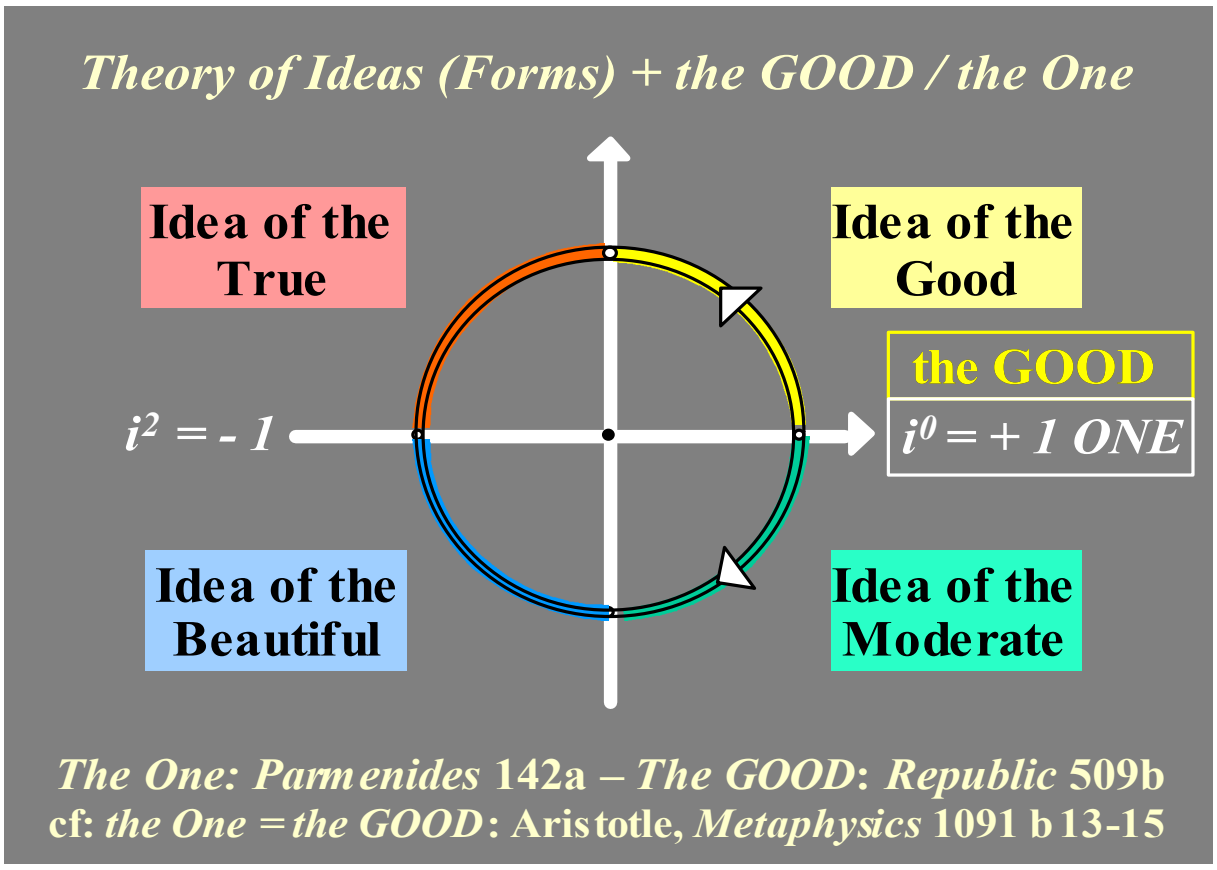

Fig. 14: Plato’s Theory of Forms (Ideas)

By the way, Plato’s ‘Theory of Forms’ corresponds analogously with his mathematical-physical cognitions. As proof may serve the tradition of Aristotle that Plato identified the GOOD with the ONE (Aristotle, *Metaphysics* 1091, b, 13-15) (Figure 14)

## Matrix and Symmetry

Filling the matrix with interdependent physical dimensions is like discovering the white areas on a map of an unknown landscape. However, in summarising it can be stated that it is possible to fill the general matrix of coherent physical dimensions with nearly all actual main physical definitions at least to the terms shown in Figure 15. Most of those definitions have a relation to Plato's texts, the rest is deduced from the modern SI-definitions and is accordingly classified.

This basic matrix (Figure 15) offers a playful approach to achieving interdependence and coherence between all physical sectors (e.g. vibrations, electricity, magnetism, optics, and even atom physics) and their specific values, as a unifying principle. In this kind of matrix, all physical definitions of all physical sectors will find an interactive correct place (Figure 17, 18) (Schweitzer. 145)

Some definitions as for example 'force', 'energy', 'performance' or 'resistance' are subject of several sectors of physics. Other dimensional positions have different physical definitions or denominations as for example the position of 'mass' is in the sector of electricity connoted with 'inductivity' which points to a deep inner relation of the addressed sectors of physics, or as another sample, 'energy', 'capacity', and 'charge' occupy the same position in this basic cyclic of dimensions (Figure 17, 18) (Schweitzer. 145). Hopefully the matrix will cause physicists to fill white spots in the matrices.

When searching the dimensional position of all known physical definitions in all fields of physics it can be observed that some of the dimensional definitions of certain fields of physics overlap with their position the igniting cycle and reach into the next upper cycle, however, there they will show up in an expected typical position.

## *Matrix of Mechanical Units*

| | | | | | |
|---|---|---|---|---|---|
| $d^{-4}$ | $d^0$ $+1$ power$_1$ | $d^4$ $d^{4+16n}$ velocity | $d^8$ gravitat. potential | $d^{12}$ resistance | $D^{16}$ power$_2$ |
| $d^{-3}$ time | $d^1$ $+i$ path energy$_1$ | $d^5$ $D^{21}$ aether | $d^9$ mass | $d^{13}$ impulse | $D^{17}$ energy$_2$ |
| $d^{-2}$ | $d^2$ $-1$ plane | $d^6$ density | $d^{10}$ | $d^{14}$ pressure | $D^{18}$ |
| $d^{-1}$ | $d^3$ $-i$ space | $d^7$ accelerat. | $d^{11}$ inertia | $d^{15}$ | $D^{19}$ |

Fig. 15: Matrix of Actual Physical Units

Basic definitions

| | | |
|---|---|---|
| meter (m) | = | $d^1$ |
| kilogram (kg) | = | $d^9$ |
| second (sec) | = | $d^{-3}$ |

| | | | |
|---|---|---|---|
| power | mkg/sec$^2$ | = | $d^1 d^9/d^{-6}$ |
| velocity | m/sec | = | $d1/d^{-3}$ |
| energy | m$^2$kg/sec$^2$ | = | $d^2 d^9/d^{-6}$ |
| aether | m$^2$kg/sec$^2$ · m/sec | = | $D^{17} d^4$ |
| density | kg/m$^3$ | = | $d^9/d^3$ |
| acceleration | m/sec$^2$ | = | $d^1/d^{-6}$ |
| grav.potential | kg/m | = | $d^9/d^1$ |
| mass | kg | = | $D^{16}/d^7$ |
| inertia | kgm$^2$ | = | $d^9 d^2$ |
| resistance | kg/sec | = | $d^9/d^{-3}$ |
| impulse | mkg/sec | = | $d^1 d^9/d^{-3}$ |
| pressure | m kg/sec$^2$ : m$^2$ | = | $d^1 d^9/d^{-6} d^2$ |
| time | m/m$^4$ | = | $d^1/d^4$ |

Fig. 16: Secured Definitions on One-Dimensional Basis

## Matrix of Dimensions - Mechanics

| | | | |
|---|---|---|---|
| $D^{16}$ $d^0$ +1<br>power<br>F<br>$mkgs^{-2}$ | $d^4$<br>velocity $m^4$<br>performance<br>v c<br>$ms^{-1}$ $m^2kgs^{-3}$ | $d^8$<br>gravity-poten.<br>energy-dose<br>j<br>$kgm^{-1}$ $m^2s^{-2}$ | $d^{12}$<br>resistance<br>mass-flow<br>*m(t)<br>$kgs^{-1}$ |
| $D^{17}$ $d^1$ +i<br>length path<br>work energy<br>l E<br>$m$ $m^2kgs^{-2}$ | $d^5$<br>kin. viskos.<br>aether $m^3kgs^{-3}$<br>V<br>$m^2s^{-1}$ $m^{-1}kgs$ | $d^9$<br>mass<br>m<br>$kg$ | $d^{13}$<br>time $m^{-3}$<br>impuls-moment<br>t P<br>$mkgs^{-1}$ $s$ |
| $D^{18}$ $d^2$ −1<br>area<br>A<br>$m^2$ | $d^6$<br>density<br>volum-stream<br>z<br>$kgm^{-3}$ $m^3s^{-1}$ | $d^{10}$<br>mass str.dens.<br>l<br>$kgm^{-2}s^{-1}$ $kgm$ | $d^{14}$<br>pressure<br>energ.density<br>p<br>$m^{-1}kgs^{-2}$ |
| $D^{19}$ $d^3$ −i<br>volume -space<br>frequency<br>R<br>$m^3$ $s^{-1}$ | $d^7$<br>acceleration<br>a<br>$ms^{-2}$ | $d^{11}$<br>inertiamom.<br>dyn. viskos.<br>B<br>$kgs^{-1}m^{-1}$ $kgm^2$ | $d^{15}$<br>$m^{-1}$ |

Fig. 17: Matrix of SI Units of Mechanics

## Matrix of Dimensions - ELT

| | | | |
|---|---|---|---|
| $D^{16}$ $d^0$ +1<br>elt.power voltage<br>permittivity<br>U e V<br>$kgm^2s^{-3}A^{-1}$ N | $D^{20}$ $d^4$<br>elt.current<br>performance<br>I P W A<br>$kgm^2s^{-3}$ $Js^{-1}$ | $D^{24}$ $d^8$<br>permeability<br>$m_0$ $Hm^{-1}$<br>$kgms^{-2}A^{-2}$ | $D^{28}$ $d^{12}$<br>elt. resistance<br>R W<br>$m^2kgs^{-3}A^{-2}$ |
| $D^{17}$ $d^1$ +i<br>charge, elt.flux<br>elt.capacity<br>q C C F J<br>$kg^{-1}m^{-2}s^4A^2$ As | $D^{21}$ $d^5$<br>aether<br>$m^2s^{-1}$ $m^{-1}kgs$ | $D^{25}$ $d^9$<br>inductivity<br>L H<br>$kgm^2s^{-2}A^{-2}$ | $D^{29}$ $d^{13}$<br>magn.flux<br>elt.resistivity<br>$Q_m$ We<br>$Kgm^2s^{-2}A^{-1}$ Vs |
| $D^{18}$ $d^2$ −1<br>elt. moment<br>current density<br>$m_e$<br>msA Cm $Am^{-2}$ | $D^{22}$ $d^6$<br>magn.moment<br>m<br>$Am^2$ | $D^{26}$ $d^{10}$<br>kgm | $D^{30}$ $d^{14}$<br>elt.charge-density<br>h<br>$Cm^{-3}$ |
| $D^{19}$ $d^3$ −i<br>magn.<br>field-strength<br>H<br>$Am^{-1}$ | $D^{23}$ $d^7$<br>$kgm^{-2}$ $ms^{-2}$ | $D^{27}$ $d^{11}$<br>magn. flux<br>density<br>B T<br>$kgs^{?2}A^{?1}$ $Wbm^{-1}$ | $D^{31}$ $d^{15}$<br>elt.flux-density<br>elt.field-strength<br>E<br>$mkgs^{-3}A^{-1}$ $Vm^{-1}$ |

Fig. 18: Matrix of SI Units of ELT – Some White Spots to be Defined

Once more it has to be registered that within this dimensional system of physical definitions all SI-basic units - not only meter, seconds, and kilogram - can be expressed by the powers of the one single unit of self motion (Schweitzer. 145).

The repeated and condensed upper cycles of dimensions with the same basic physical definitions let the slight hope arise that by this reviewable dimensional system of physical definitions a superordinate system as unifying concept of different sectors of physics could be thinkable.

The cyclical system of physical dimensions evocates the most interesting hypothesis that all of those dimensions in this cyclical system are always interconnected because all of them are based on the basic unit of self movement and any unit's size may influence the others to a certain degree at any stage or any process. It is the standpoint of the observer, his viewing direction, and his intention which let him register the selected and expected influences.

In any case, the here deduced matrix of physical dimensions as potencies on a complex base might as humble result serve to check physical formulas corning their correctness or detecting the dimensions of single terms of such equations just by adding or subtracting the respective exponents of their dimensional values. Especially students might take pleasure in comparing this Platonic system of interdependent units with the SI-catalogue of physical units.

## Aether

A remarkable result of the matrix is regarding the utmost speed, the speed of light, and the consistency with Einstein's $E = mc^2$ or $D^{17} = d^9 \cdot d^8$. Furthermore, in light of the classification of physical units the serious question arises, what the matter between energy and mass, $E \cdot c$ or $d^1 \cdot d^4 = d^5$, or in the next cyclic $D^{17} \cdot d^4 = d^{21}$ etc., could be? This medium between energy and mass might as working hypothesis be regarded as 'aether' (Figure 15, 16). For Plato 'aether' is just the clearest kind of the element air (*Timaeus* 58d). In the late 19th century, theories were discussed regarding the aether as a luminiferous (light-bearing) medium responsible for the propagation of light. Einstein ranged between refusal and acceptance of the aether theories (Kostro). Today, the majority of scientists refute the existence of an aether medium, although a number of physicists promote the belief in such medium.

## Plato's Unifying View of the World

After this short unifying view of physics and ethics, derived from Plato's circular sequence of dimensions and his dimensional sequence of elements (*Timaeus* 32d), we have in a kind of dihairesis consider Plato's way of thinking, which he explained at the sample of a circle by the graded valences of intellectual stages of reflection (*Letters* VII 342a-344c) (c.f. *Laws* X 898a; *Timaeus* 47b,c, 90d). Also Plato's *Analogy of the Line* understood as a divided outline of circle surliness his circular thinking. Thus Plato found by his cyclical system of powers on a complex base a unifying view of the world with its coherent sub-worlds. It was a philosophical challenge to dissect Plato's unifying structure of metaphysics (Schweitzer. *The Atlantis Irony*).

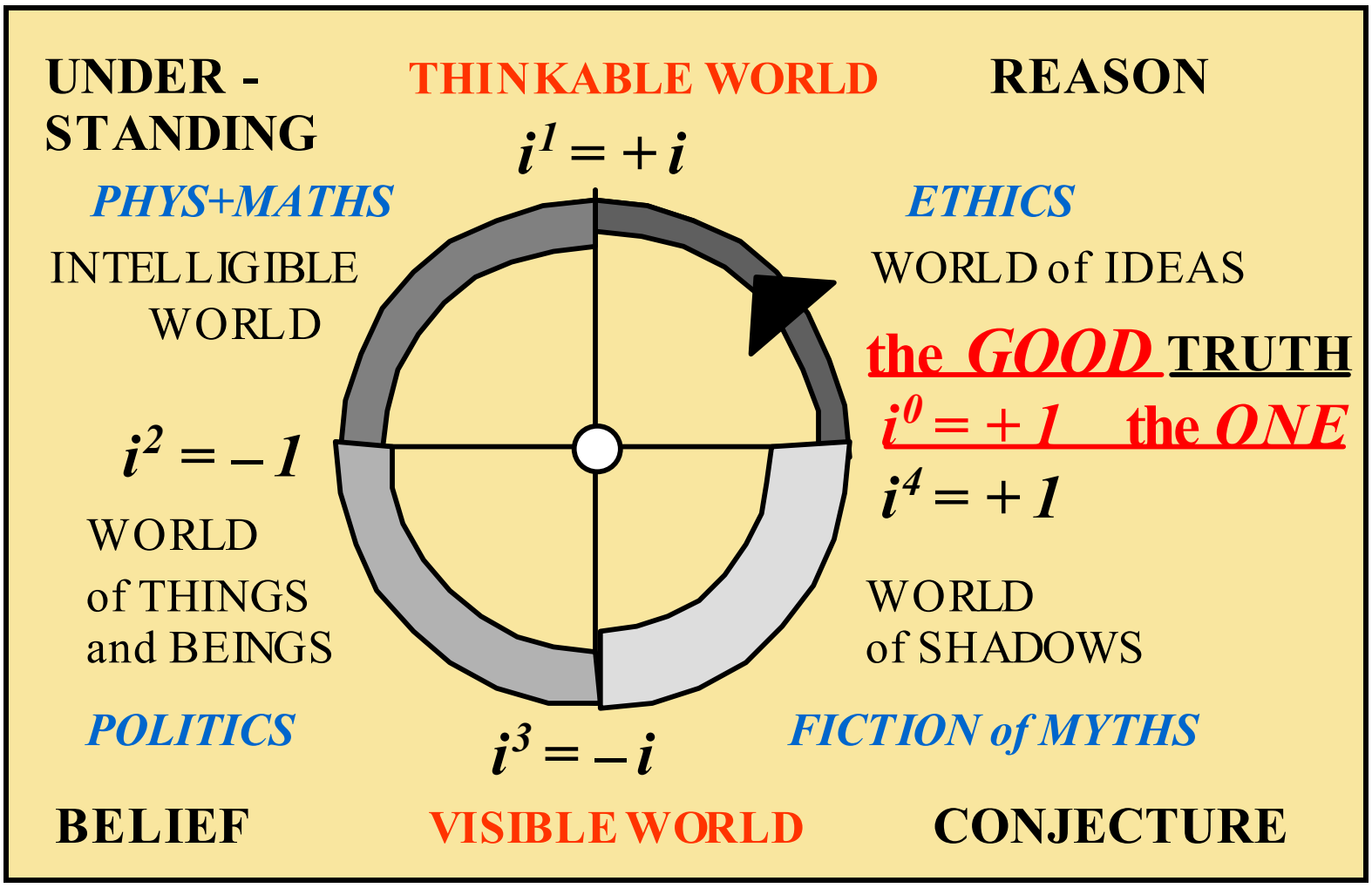

Fig. 17: Plato's Worlds and Thinking Order

## Plato's Reason for Encoding his Cognitions

Plato argues that during a serious dialectical process, the cryptic logos will talk to those who are qualified, or remain silent with those who are not qualified (*Phaedrus* 275d–276a). Plato worried that his teachings might fall into the wrong hands (*Republic* VII 536a–b, and VII 537c–539d) (*Phaedrus* 275e) (*Letters* VII 341e). He clearly says that he is hiding the fundamental principle in cryptic expressions so that incapable people, who might laugh at it, could not possess it (*Letters* II 312d–e and II 314a). However, he also recognized the danger that a broad general knowledge of his teachings could endanger the education of qualified persons who are necessary to perpetuate the republic and the constitution (*Republic* VII 536a–b). Plato offered his posterity the possibility to tackle his mirrored words in accordance with everyone's individual skills and abilities. So everybody has the chance to successfully unravel his coded messages according up to his mental level and nobody must in any way feel discriminated. On the other hand, Plato prearranged that only a few persons would be able to find the hidden truth through logic (*Republic* VII 532d–e).

## Conclusion

Plato had definitely access to the field of complex numbers and complex geometry, but unfortunately, no one was able to follow him. It was not until the Renaissance when complex figures were discovered again, and it took more than 2000 years until Gauss formulated the principles of complex mathematics. For Plato, his general insight into complex mathematics seemed to have been a reasonable basis for his understanding of physics. And he used his respective physical cognitions as analogon for the order of his ideas and finally as proof for his ethical creed, the cognition of the Good. He used the cycle of powers of complex values also for explaining the way of thinking. Subordinating the classification of his philosophy to this circular system offered him the differentiated view of the world and its subdivisions with their respective intellectual requirements. Insofar this circular dimensional system meant to him a unifying view of the factual world, natural sciences, humanities and ethics.

In the natural scientific respect, the here updated Platonic unifying system of physical dimensions shows clearly the interrelation of <u>all</u> physical dimensions based on a primordial power of self-motion. The here demonstrated interconnections at Plato's physical dimensiology are de facto reviewable and can be demonstrated by accordingly checking all physical formulas. Additionally there is a lot of space for detecting new physical correlations.

It was a philosophical challenge to at least partly dissect Plato's mostly gradually ironically coded unifying structure of metaphysics by thinking outside the box. This modest attempt to adopt Plato's physical insights to today's traditional physics is the first speculative access to the trial of understanding Plato's physical dimensiology, at least to a certain degree. The result of this research is of a certain interest for History of Science because of the knowledge about fundamental insights in complex mathematics and mechanical physics Plato had already 2350 years ago.